\documentclass{sig-alternate}

\usepackage[table]{xcolor}

\usepackage{tikz}
\newcommand*\circled[1]{\tikz[baseline=(char.base)]{
            \node[shape=circle,draw,inner sep=2pt] (char) {#1};}}
\usepackage{color}
\usepackage{framed}
\usepackage{comment}
\usepackage{todonotes}
\usepackage{soul}
\usepackage{url}
\usepackage[caption=false]{subfig}
\usepackage[hidelinks]{hyperref}
\usepackage{mathtools}

\newcommand{\ie}{{\em i.e.,\/ }}
\newcommand{\eg}{{\em e.g.,\/ }}

\newcommand{\etal}{{\em et al. \/}}

\makeatletter
\def\@copyrightspace{\relax}
\makeatother

\begin{document}
\clubpenalty=10000 
\widowpenalty = 10000

\title{LIRA: A Location Independent Routing Layer \\ based on Source-Provided Ephemeral Names
}

\numberofauthors{1} 
\author{
\alignauthor
Ioannis Psaras, Konstantinos V. Katsaros, Lorenzo Saino, George Pavlou \\
       \affaddr{Dept. of Electrical \& Electronic Engineering, University College London}\\
       \affaddr{WC1E 7JE, Torrington Place, London, UK}\\
       \email{\{i.psaras, k.katsaros, l.saino, g.pavlou\}@ucl.ac.uk}
}

\maketitle

\begin{abstract}

We identify the obstacles hindering the deployment of Information Centric Networking (ICN) and the shift from the current IP architecture. In particular, we argue that scalability of name resolution and the lack of control of content access from content providers are two important barriers that keep ICN away from deployment.
We design solutions to incentivise ICN deployment and present a new network architecture that incorporates an extra layer in the protocol stack (the \textit{Location Independent Routing Layer}, LIRA) to integrate location-independent content delivery.
According to our design, content names need not (and should not) be \textit{permanent}, but rather should be \textit{ephemeral}.
Resolution of non-permanent names requires the involvement of content providers, enabling desirable features such as request logging and cache purging, while avoiding the need for the deployment of a new name resolution infrastructure. Our results show that with half of the network's nodes operating under the LIRA framework, we can get the full gain of the ICN mode of operation.

\end{abstract}


\makeatletter{}
\section{Introduction}

Network routing based on content identifiers has recently become a topic of extensive discussion, due to the benefits that could be provided by a location-independent data distribution network \cite{surveyICN}, more commonly referred to as an Information-Centric Network (ICN).
For instance, the ICN \textit{request-response} mode of operation alleviates client mobility issues \cite{garethMobility} and natively supports interdomain multicast \cite{SarelaBFInfocom}. 
Furthermore, content security (as opposed to channel security) is inherently supported by transmitting signed copies of content \cite{ccn}.
This in turn allows for in-network caching, which can transform the Internet into a native content distribution network \cite{cachingSurvey}.
Finally, as shown recently \cite{inrpp-hotnets-ipsaras}, the ICN paradigm can bring benefits also at the transport layer, where caches can be exploited to alleviate congestion.

On the other hand, enormous effort has been spent to de-ossify the end-to-end Internet transmission model to enable new functionalities.
Examples include IP Multicast and Anycast and supporting IP mobility at the network layer \cite{multicast}, \cite{ip-anycast-sigcomm}, \cite{mobileip}.
However, the difficulties of deploying those solutions at large scale led to the design of application-layer solutions such as overlay caching instead of native, in-network caching, overlay indirection techniques \cite{stoica-i3}, \cite{http-narrow-waist}, \cite{multicache}, DNSSEC and IPSEC to enhance security, just to name a few.
Even though these solutions have the potential to enable new services (or applications), they appear inferior compared to an ICN mode of operation, as they cannot natively support security, mobility, in-network caching and multicast: in all cases the in-network forwarding entities are forced to operate on the five-tuple \texttt{<sourceIP, destinationIP, sourcePort,  destinationPort, protocol>} being, therefore, completely content-agnostic.

Arguably, the ICN paradigm has the potential to deal with the Internet's most daunting problems in a native manner.
To reach this point, however, a new architecture based on core ICN principles will have to be deployed over the current IP Internet architecture, clearly, a rather challenging task.

In this paper, we identify two main obstacles that hinder the deployment of ICN on top of the current Internet.
These are: \textit{i)} scalability of name resolution, a core networking problem \cite{caesar-ancs14} and \textit{ii)} content provider-controlled access to content, a business model problem, which however, is deeply integrated into the core networking principles of today's Internet and therefore, affects the design of any new architecture.
Content access control here is linked to content access logging and the transmission of content transparently to the content provider from in-network caches. 
We discuss each of these two challenges in more detail next.
Based on these considerations, in this paper, we propose a fully backward-compatible and incrementally-deployable ICN-oriented architecture that meets scalability concerns, but at the same time takes into account the business requirements of the main Internet market players.

\subsection{Name resolution scalability}

Two main schools of thought have emerged in the ICN-related literature regarding name resolution and name-to-location mapping.
The first one, mainly adopted by the original CCN/NDN proposal \cite{ccn}, advocates the hop-by-hop resolution of \textit{requests} or \textit{Interests} at the data plane. 
Effectively, name resolution is coupled with name-based forwarding with each Interest packet being locally resolved to the next (router) hop. This approach has the advantage of locally making forwarding decisions, but on the downside, huge volumes of state need to be maintained in (manually-set) FIB tables \cite{reality}.
CCN/NDN routers effectively have to keep state \textit{per packet}, an issue traditionally considered as an implementation challenge~\cite{tsiloInfocom}.\footnote{See also \cite{web-icn-oran} for an elaborate discussion on issues related to web transfers where the current NDN model is not sufficient.}
To deal with the scalability problems \cite{networking15} of the original proposal and the huge state that needs to be kept at all routers, recent developments in the NDN space have proposed an NDN-based DNS system, dubbed NDNS \cite{afanasyev-phd}, as well as the involvement of content providers to help in the name-resolution process \cite{snamp-gi15}.

The second school of thought decouples name-resolution from name-based routing by using a separate name-resolution system, similar in nature to DNS (\textit{e.g.}, \cite{netinf-comcom}, \cite{curling}, \cite{tyson-juno}, \cite{pursuitnrs}). Although this approach avoids pushing excessive state to router forwarding tables, it requires the deployment of new infrastructure by operators. For instance, as shown in \cite{networking15}, the support of the DONA~\cite{dona} architecture at tier-1 Autonomous Systems (ASes) requires the deployment of small-to-medium size data centres to support name resolution. Such, extra infrastructure built in from scratch has the obvious downsides of huge investment requirements, as well as the shift challenge to this new mode of operation.

Moreover, focusing on the practical deployment of ICN, the full cycle of the name resolution process still remains unclear. 
Name resolution and data delivery mechanisms often build on the implicit assumption that content names or identifiers are already available to the end users, prior the aforementioned coupled or decoupled resolution steps. 
Obviously, developing a mechanism for the retrieval and delivery of content names to the end users raises concerns regarding both scalability aspects related to the enormous size of the namespace, and compatibility issues with respect to both application and network layer interfaces. 

We also note that the requirements of today's dynamic and interactive applications would not be served adequately by fully transparent in-network caching driven solely from search engine content name results. We discuss and evaluate these concerns later.

\subsection{Content access control}

Content Providers (CPs) and CDNs require, for commercial and regulatory reasons, full control over the content requested and transferred.
This has been largely overlooked by research efforts in the ICN area, which have mainly focused on naming schemes and name resolution systems to address scalability issues.
For instance, the consensus around {opaque} and {permanent} content names ignores the fact that content can be served from ISP-operated in-network caches, transparently to the CP or CDN.
\textit{``Pay per click"} business models, however, would face significant limitations from this design choice in an ICN setting, that would practically prevent CPs and CDNs from billing their customers. 
Alternative approaches based on ISP-CDN collaborations to log content requests cannot but be unrealistic: DNSs can keep track of requested content and could possibly report back to the relevant CPs/CDNs. This, however, would mean that SLAs should be in place between \textit{all ISPs and all CPs/CDNs at a global scale}, a rather unrealistic assumption.

At the same time, transparent in-network caching mechanisms would typically allow only limited control over the content delivered to clients. That is, coarse grained TTL-based mechanisms would be the only means for CPs/CDNs to manipulate updated content, leading either to the delivery of stale content, or the unnecessary delivery from the CP. That said, active cache purging is another requirement that calls for control of content from CPs and CDNs.

Although content access control might sound as a trivial implementation or a business model issue, we argue that it might well hinder the engagement of CPs and CDNs from the adoption of ICN. Summarising, we argue that these concerns of: \textit{i)} scalability and incremental deployment support of a name-oriented architecture, and \textit{ii)} exclusive content access control at the CP side with simultaneous support for transparent in-network caching 
 have been overlooked by the community so far. As a consequence, the full potential of an ICN mode of operation has not been exploited in full yet, making the adoption and deployment of the ICN paradigm an unrealistic target.

\subsection{Contributions}

Although clean slate research has revealed many of the benefits that ICN can bring, we argue that deployability has to be put at the forefront of any ICN design, rather than being treated as an afterthought.
We address the deployability concerns discussed above by introducing a novel information-focused network architecture, which overcomes scalability concerns and is fully backward compatible with the current IP architecture.

Our proposed architecture first introduces a name resolution process tailored to carefully manage information exposure \eg enabling content access logging (Section \ref{cc::cpbnm}). This name resolution process is combined with a new naming scheme, which builds on the notion of \textit{ephemeral names} (Section \ref{cc:en}).
\textit{Name resolution is controlled by content providers} based on a fully backward-compatible mechanism that supports in-network caching and the direct control of ephemeral names' lifetime, thus facilitating content access logging and active purging of stale cached data.
The proposed mechanism completes the full cycle of the name resolution process, delivering content names to clients, without imposing any requirement for additional mechanisms.

In-network caching, name-based routing and support for network-layer multicast are all integrated in the \textit{Location-Independent Routing Layer} (LIRA), an extra layer in the protocol stack placed at ``level 3.5" of the protocol stack, above the IP and below the transport-layer (Section~\ref{cc:cl}). 
LIRA ``absorbs" the location-independence nature of ICN, leaving the network layer to operate based on IP addresses.
Resolution of content names does not rely on large volumes of FIB table entries, and routing takes place based on a hybrid of IP addresses (at the IP layer) and location-independent transient content names (at the LIRA layer) (Section \ref{rsn-details}).
Our design does not require blanket adoption in order to realise the benefits of ICN. 
Instead, ISPs can incrementally deploy LIRA nodes with little investment. 
Furthermore, the fact that routing is (in the worst case) based on IP addresses guarantees full backward compatibility with the current Internet architecture.
Our results show that even with a subset of nodes upgraded to support LIRA functionality, our design achieves considerable performance gains (Section \ref{eval}).

\makeatletter{}

\section{Concepts and Components}\label{cc}

\subsection{Content provider-assisted name resolution}\label{cc::cpbnm}

In order to deal with the scalability concerns raised above, we design a name resolution scheme which involves the content provider and does not require extra name-based resolution machinery (\eg \cite{curling}, \cite{tyson-juno}, \cite{netinf-comcom}, \cite{copss-mayutan}), or manually-set, bloated FIB tables (\eg \cite{ccn}, \cite{conet}). In particular, any user will have to consult the CP (or CDN) and ``ask" for the name/\texttt{contentID} before any content transfer can start (see next section for details on the \texttt{contentID}).
Users reach the content provider based on the standard procedure of the current Internet, that is, based on URLs, DNS resolution and IP addresses. This first part of the resolution (\ie reaching the CP to get the \texttt{contentID}) is based on IP addresses and is location-dependent. We note that users do not get the whole of the chunk from the CP (but only the \texttt{contentID}), which can be served from any other cache in the network. In this way, we realise \textit{semi-transparent} in-network content caching, which we argue is in the best interests of both CP/CDNs and ISPs alike. As discussed later on in this section, the second part of the name resolution, which also leads to the content transfer itself is location-independent, according to the philosophy of ICN. Summarising, the \textit{``content provider-controlled name resolution procedure"} introduced here is fully backward compatible and does not require extra investment from ISPs, or CPs/CDNs.

\subsection{Ephemeral Names}\label{cc:en}

To provide full content access control to CPs, we introduce the concept of \textit{ephemeral names}, which are used for location-independent content delivery. Our primary motivation behind the introduction of \textit{ephemeral names} is to avoid dissemination of the name/\texttt{cID} of a content to other users, as this could potentially lead to accessing the content from in-network caches, transparently to the CP/CDN.
This section explains the structure, usage and design principles of these names.

\subsubsection{Name Structure}

\begin{figure}[h!]
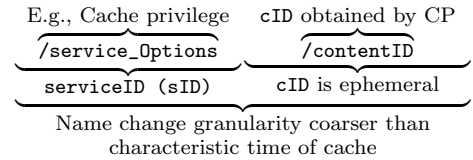

\begin{centering}
\small
$
\underbrace{\underbrace{ \overbrace{\hbox{\texttt{/service\_Options}}}^{\parbox{3cm}{\centering E.g., Cache privilege}}}_{\hbox{\texttt{serviceID (sID)}}} \underbrace{\overbrace{\hbox{\texttt{/contentID}}}^{\parbox{3cm}{\centering \texttt{cID} obtained by CP }}}_{\parbox{3cm}{\centering \texttt{cID} is ephemeral}}}_{\parbox{6cm}{\centering Name change granularity coarser than characteristic time of cache}}
$
\caption{Ephemeral Names}
\label{ephemeral-names}
\end{centering}
\end{figure}

The LIRA architecture uses flat names composed of two parts (see Fig.~\ref{ephemeral-names}). The main part of the naming structure, the \texttt{contentID}, or \texttt{cID} reflects the name of the content itself and is based on the premise of \textit{ephemeral or transient names}. According to this concept, content providers choose arbitrary strings and assign them to the content they host. The names are flat, in the sense that they bear no structure related to routing (\eg aggregation); however, CPs may impose structures related to the internal organisation of their content. Ephemeral names should be unique to guarantee collision-free name resolution, which can be easily achieved with the use of arbitrary hashes. The names are self-certifying and ``expire" after some time interval.\footnote{A few randomly set padding bits can be used in each named chunk to preserve both the self-certifying and the ephemeral character of names, without inflating the chunk size.}
This transitioning interval should be coarser than the time needed to support in-network caching and multicast (\eg names should not change on a per-request basis) - see Section~\ref{transitioning-interval} for details.

\normalsize

The second part of the \textit{ephemeral name}, the \texttt{serviceOptions}, can be used to realise preferential treatment of content. Although the use of this part of the name is not necessary in our architecture, and is not necessarily of ephemeral nature, we believe that it can help in the caching and scheduling process. For instance, the \texttt{serviceOptions} part can be used to flag content that should or should not be cached. We leave such investigations for future work.

\subsubsection{Incentives and Disincentives for Adoption}

In case of permanent names, search engines would operate based on names (similarly to today's operation based on URLs). This operation is clearly not in the best interests of CPs/CDNs given the \textit{``pay per click"} models in use today and transparent in-network caches used in ICN. 

Transient names dis-incentivise search engines from disseminating \texttt{cIDs}, but at the same time allow for both access logging at the CP/CDN and transparent in-network caching. One might claim that search engines would prefer to provide the \texttt{cID} directly to users, as this would lead to faster content access (\ie users would not need the extra RTT to travel to the CP/CDN to get the \texttt{cID}). However, given (i) the transient character of names, and (ii) the delivery of bundles of \texttt{cID}s by CPs (see Section~\ref{transitioning-interval}), this would require search engines to devise mechanisms for retrieving and disseminating \texttt{cID}s each time they change, only to save a single RTT in each bundle. This limits the incentives of search engines to provide \texttt{cID}s without the consent of CPs/CDNs.

Moreover, and most importantly, ephemeral names allow CPs/CDNs to actively control the cached content served to their clients  \eg by changing the \texttt{cID}s of content chunks existing cached copies get practically invalidated. This is an important feature of the proposed approach, which cannot be supported in alternative proposals (\eg \cite{ccn,netinf,conet}).

\subsubsection{Transitioning Interval}\label{transitioning-interval}

The combination of the name resolution at the CP, together with the ephemeral nature of content names supports a number of desirable features. First and foremost, name resolution is under the control of the CP, enabling access logging. Secondly, versioning of updated content and purging of old content from in-network caches is also under the control of the CP. 
Although TTL-like techniques, such as the CCN staleness option, can support content updating, it is not easy to set such values given today's interactive applications. Setting TTL values for individual content items (\eg~\cite{netifncachecontrol}) would always face the tradeoff of short TTLs resulting in unnecessary delivery from the content provider, while longer TTLs would result in delivering outdated content. Using ephemeral names, cached content can instead be actively invalidated when needed. 

Along the same lines, the transitioning interval of ephemeral content names is an issue that requires further attention and is related, among others, to the popularity of the content as well as the size of content chunks. Frequent change of the name can result in suboptimal performance, since each change purges the content in caches. We deal with this tradeoff by setting the transitioning interval of content names to a value inversely proportional to the popularity of the content itself. Popularity is measured by the CP and can be based on the number of requests for the content in question, per some time interval. Although more sophisticated settings can be found, with this simple setting for the transitioning interval we avoid changing the \texttt{cID} of rarely accessed content too frequently, and we also avoid leaving the \texttt{cID} of popular content the same for too long. 
Finally, to alleviate the need to travel to the CP for every chunk request, we assume that upon each request for a content item, the CP sends back to the client the ``up-to-date" ephemeral names of the next few subsequent chunks, that is, not only the name of the immediately following one. The number of subsequent \texttt{cID}s sent by the CP to the client is left for future investigation.

\subsection{LIRA: Location-Independent Routing Layer}\label{cc:cl}

Adding extra functionality, or altering completely the operation of \textit{existing} core network protocols can prove difficult to be done incrementally (\eg IPv6) and ``flag-days" are not an option for incorporating new components at a global scale. For these reasons, we propose \textit{addition} instead of \textit{replacement} of an extra layer to the protocol stack, which we call \textit{Location-Independent Routing Layer} (LIRA). LIRA sits on top of the network (IP) layer and below the transport layer. It operates based on \textit{ephemeral names} and integrates all the required functionality to realise \textit{location independence}, taking advantage of \textit{information centricity} and its well-known gains \cite{surveyICN}.

Although recent studies have proposed HTTP as the layer that can integrate information or content centricity \cite{http-narrow-waist}, here we argue that in order for in-network caching and multicast to be smoothly incorporated in the new ecosystem, any information-centric operation needs to be \textit{below the transport layer}. Otherwise, the transport protocol can merely connect two specific endpoints cancelling any notion of location-independent content transfer. Instead, breaking the end-to-end transmission model below the transport layer allows to leverage (ICN enabled) in-network caching, both in terms of native multi-source routing and localised congestion control \cite{inrpp-hotnets-ipsaras}, going far beyond traditional IP Multicast or Anycast mechanisms.

LIRA is implemented in just a small subset of nodes (see Section~\ref{lira-nodes}), which can be transparently planted in the network, and it manages incoming and outgoing content based on their names. The main name management functionality is implemented in a routing table, which we call \textit{Content Forwarding Information Base} (C-FIB) (Section~\ref{rsn-table}).

A similar notion to the LIRA layer has been proposed in the past in \cite{triad}, but in a totally different context, addressing the exhaustion of IPv4 addresses. The evolution of NAT boxes (together with the painfully slow incremental deployment of IPv6) has dealt with this problem and hence, the related efforts became obsolete.

\subsection{Content Forwarding Information Base}\label{rsn-table}

The Content Forwarding Information Base (C-FIB) table keeps track of recently requested and served content (in terms of \texttt{cID}s) and maintains forwarding information used for the delivery of those content items, providing also support for in-network caching and multicast. Upon subsequent request(s) for a content already in the C-FIB table, LIRA is redirecting requests towards the direction where the content has been sent, or served from, similarly in principle to breadcrumbs routing \cite{breadcrumbs}.

We note that \textit{the C-FIB table essentially acts as a cache} for \texttt{cID}s served recently through this router (somewhat similarly to \cite{scan} and \cite{conet-comnet}). However, C-FIB table entries are not permanent, as in CCN's FIB, but rather are assisting in location-independent content delivery from neighbouring nodes (see Section~\ref{rsn-details} for details on the C-FIB structure).

The typical structure of the C-FIB table is illustrated in Table~\ref{table:r1routing-table}. The table maintains one entry per content chunk. The following information is maintained for each entry: \textit{i)} \texttt{cID}, the content identifier of the chunk, \textit{ii)} $if_{I}$, the \textit{incoming interface} \textit{i.e.}, the index of the interface from which the content is received, that is, the content source indicated by the DNS, \textit{iii)} $if_{O}$, the \textit{outgoing interface} \ie the index(es) of the interface(s) towards which the content is currently being forwarded, \textit{iv)} $if_{TI}$, the \textit{temporary interface}, \ie the index(es) of the interface(s) where the content has been forwarded, \textit{v)} $mIP$, the \textit{multicast IP} field that holds IP addresses of clients participating in a multicast session.
Note that interface entries in the C-FIB table denote real interfaces (\ie directions towards which requests/content should be forwarded) and not IP addresses of sources/destinations (apart from the multicast IP field). By doing so we realise the \textit{location independence} property of ICN in LIRA.

\subsection{LIRA Nodes}\label{lira-nodes}
The LIRA node structure is the main component of the proposed architecture, which integrates information centricity. LIRA nodes implement the LIRA layer with its C-FIB table discussed above in order to realise named content management and subsequently location independence. LIRA nodes also include caches that temporarily store named content chunks (\ie in-network caching). Although by default all LIRA nodes include both the C-FIB table and content caches, we also evaluate (in Section~\ref{eval}) the case of ``lighter" LIRA nodes, where, based on node centrality metrics and to facilitate incremental deployment, some nodes implement the C-FIB table and some others implement caches.

Our design does not require all nodes of a domain to become LIRA nodes and it is operational regardless of this. Being always based on IP, nodes fall back to normal IP operation and route towards the direction indicated by location-based addresses. Note that all routers maintain the default IP-based FIB table. Therefore, incompatibility issues or requirements for simultaneous shift to ICN operation do not exist. As we show later on in the evaluation section, an average of 50\% of nodes within a domain can provide considerable performance gain. Careful network planning (\eg depending on topological issues) and incremental upgrade of normal routers to LIRA nodes gives a major advantage to the proposed architecture in terms of deployability compared to other ICN architectures.

\makeatletter{}\section{Overview of Main Operations}\label{rsn-details}

We proceed with the description of the name resolution and content delivery process, illustrated in Fig.~\ref{fig:summary}. We then give details of the entries of the \textit{Content Forwarding Information Base} (C-FIB) table during the content delivery process. For this purpose we use the network topology presented in Fig.~\ref{example}.
Tables \ref{table:r1routing-table} and \ref{table:r2routing-table} are also used to present the entries of the C-FIB table(s) for a sequence of important events taking place in our example scenarios (denoted with timestamp $t_{i}$).

\subsection{Name Resolution and Content Delivery}\label{resolution}

The name resolution process is initiated through existing protocols (\ie DNS and HTTP) to guarantee backwards compatibility and facilitate adoption of ICN.

\begin{figure} [!t]
\begin{center}
	\includegraphics[trim= 0 10 0 0, clip=true, width=3in]{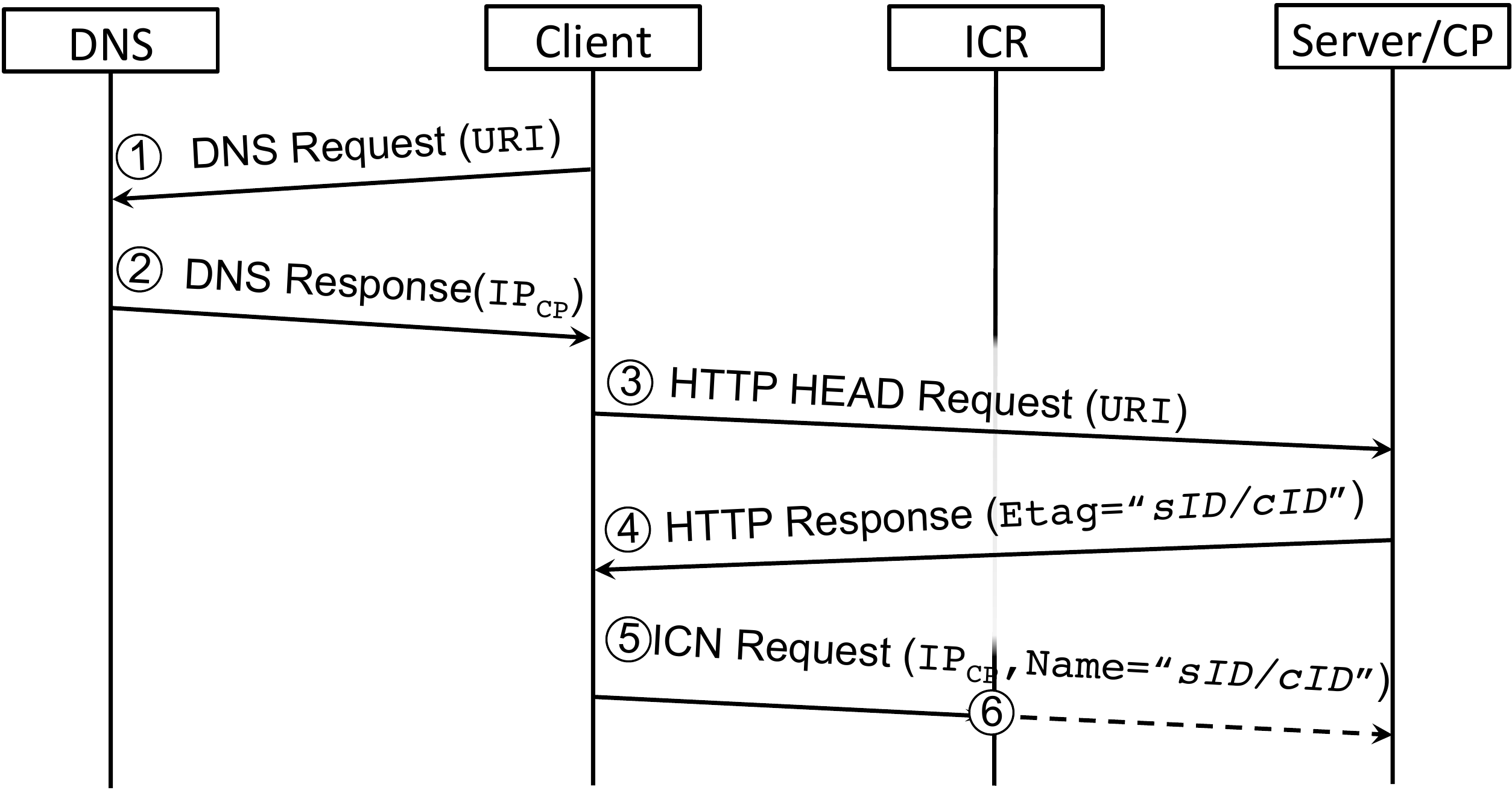}\vspace{-0.3cm}
	\caption{Name resolution and content delivery}
	\label{fig:summary}    
\end{center}
\end{figure}

\begin{figure} [!t]
\begin{center}
	\includegraphics[trim= 0 0 0 0, clip=true, width=3in]{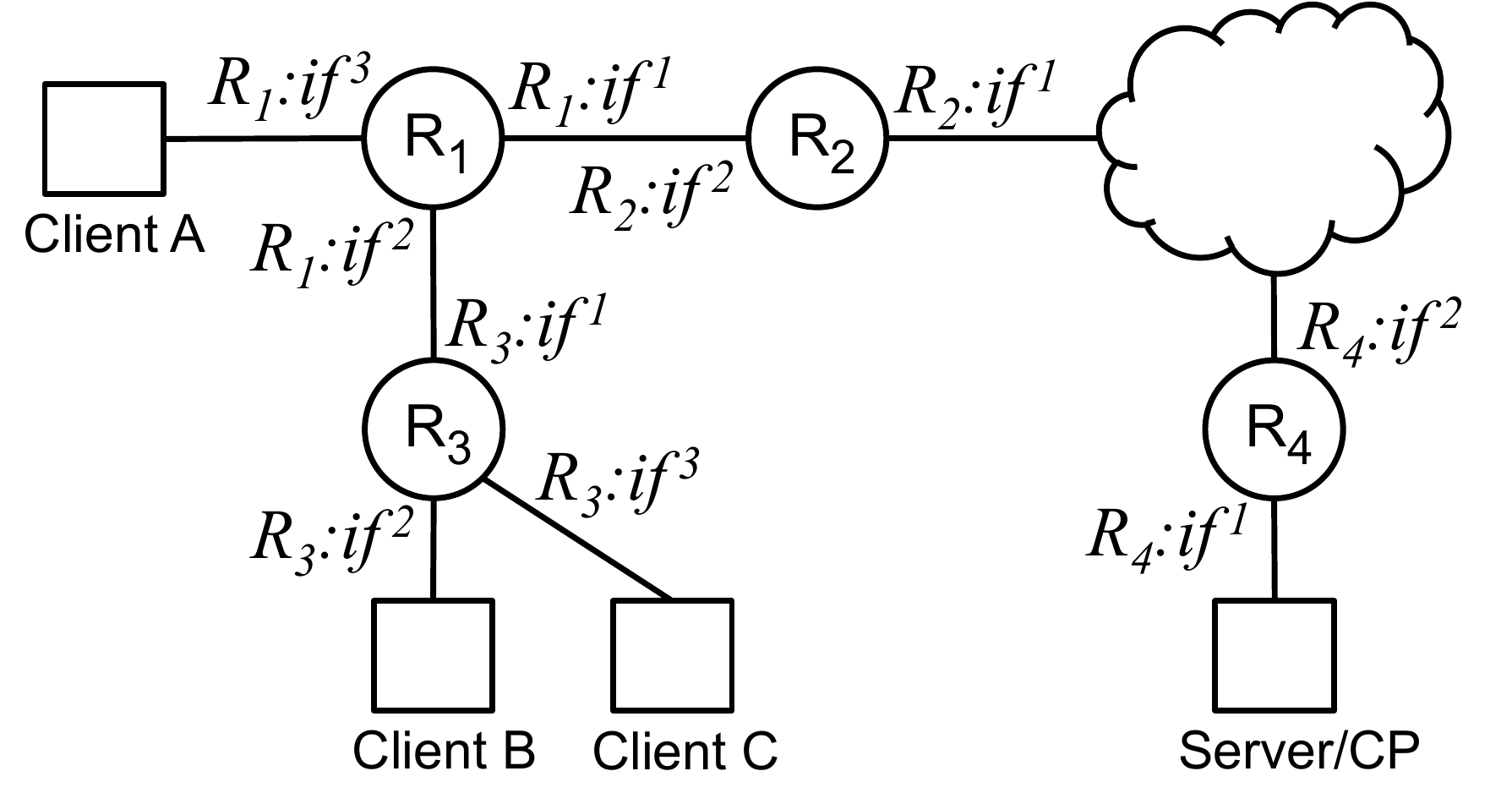}\vspace{-0.3cm}
	\caption{Example topology: labels $R_{i}:if^{j}$ denote the index $j$ of each router's ($R_{i}$) interface.}
	\label{example}    
\end{center}
\end{figure}

\circled{1} As a first step (Fig.~\ref{fig:summary}) and identically to what is happening today, users resolve URLs
through a request to the DNS.
\circled{2} The DNS responds with the IP address of the content provider.
\circled{3} The user generates an HTTP HEAD request \cite{rfc2616} at the application layer. At this stage, routing is location-dependent and is based on the IP address indicated by the DNS. At the content layer, the request is asking for the \texttt{cID}.
\circled{4} The CP sends back an HTTP response packet containing the up-to-date name, \ie \texttt{cID}, of the requested content in the ETag field of the HTTP response header \cite{rfc2616}.\footnote{It is noted that the use of the ETag field for name resolution does not change the semantics of the field as it is intended to describe the content to be delivered and/or cached. Moreover, no restrictions apply to the format of this field allowing the realisation of ephemeral names.} The destination IP address of that packet is that of the requesting client. This packet can be piggybacked with data to avoid an extra RTT between the client and the CP. In this case, however, given that requests are sent per chunk, we cannot take advantage of in-network caching. This option can be considered in special cases (\eg when a client is close to the CP and chances of finding the content cached are slim).
\circled{5} The client issues a request for the first chunk of the content object (\eg client A in the example of Fig.~\ref{example}). The request includes the IP address of the CP at the IP layer and the \texttt{cID} of the chunk at the LIRA layer.

\begin{table}[!t]
\begin{center}
\rowcolors{1}{white}{gray}

\begin{tabular}{l |*{5}{c}}
$R_1$              & $cID$ & $if_I$ & $if_O$ & $if_{TI}$ & $mIP$ \\
\hline
$t_1$	 & $x_1$ & 1 & 3 & - & -  \\
$t_2$          & $x_1$ & 1 & - & 3 & -   \\
$t_4$     	  & $x_1$ & 1 & 2 & 3 & -  \\
$t_5$     	  & $x_1$ & 1 & - & 2, 3 & -  \\
\hline
\hline
$t_7$     	  & $x_2$ & 1 & 3 & - & -  \\
$t_9$     	  & $x_2$ & 1 & 2, 3 & - & B's IP  \\
$t_{10}$     & $x_2$ & 1 & - & 2, 3 & -  \\
\end{tabular}
\caption{Routing Table at $R_1$ - Fig.~\ref{example}}
\label{table:r1routing-table}
\end{center}
\end{table}

\begin{table}[!t]
\begin{center}
\rowcolors{1}{white}{gray}

\begin{tabular}{l |*{5}{c}}
$R_3$              & $cID$ & $if_I$ & $if_O$ & $if_{TI}$ & $mIP$ \\
\hline
$t_3$	 & $x_1$ & 1 & 2 & - & -  \\
$t_6$          & $x_1$ & 1 & - & 2 & -  \\
\hline
\hline
$t_8$          & $x_2$ & 1 & 2 & - & -  \\
$t_{10}$     & $x_2$ & 1 & - & 2 & -  \\
\end{tabular}
\caption{Routing Table at $R_3$ - Fig.~\ref{example}}
\label{table:r2routing-table}
\end{center}
\end{table}

\circled{6} LIRA nodes along the path check the \texttt{cID} included in the request\footnote{We rely on the Protocol field of the IPv4 header (or the ``Next header" in the IPv6 header) to enable LIRA nodes to identify those IP packets that can be handled by the LIRA layer \ie containing an ICN content name.} against the entries of their C-FIB table. If an entry for the \texttt{cID} exists, then they forward according to this entry. If not, they forward according to the IP address. The IP address points to the CP, hence, content can always be resolved according to that in the worst case, \eg in case of LIRA-incompatible nodes or domains.

At this point, assuming the content is not locally cached (see Section~\ref{caching} for details on in-network caching), the request is forwarded towards the CP. The index of the network interface used to forward the request is marked as the $if_I$ for this content chunk (\ie interface 1 - see time $t_1$ in Table~\ref{table:r1routing-table}). At the same time, the index of network interface from which the request was received is marked as an output interface (interface 3 in our example). The content chunk is then sent back from the CP (or any other cache further down the path). During the data transfer no change is made in the C-FIB table entries of intermediate LIRA nodes (time $t_1$). When the chunk transfer completes, which is denoted by an End of Chunk (EoC) field, the intermediate LIRA nodes change their C-FIB entries for this \texttt{cID} by marking the interfaces through which they forwarded the data (\ie $if_{O}$) as $if_{TI}$ (\textit{temporary interface}) - interface 3 is moved to $if_{TI}$ at $t_2$ in Table~\ref{table:r1routing-table}. This is done since the content can possibly be delivered from there too (\ie the content has possibly been cached towards this direction).
When the client sees the EoC field/bit set, it forwards the next request towards the original CP (similarly to the initial request - step \circled{3} above) in order to obtain the \texttt{cID} of the next chunk.

\subsection{In-Network Caching}\label{caching}

LIRA nodes by default support in-network caching. In the simplest case, on-path in-network caching is supported by simply performing a lookup of the \texttt{cID} of a request message, at the local cache index. In case the requested content chunk is cached locally, the corresponding data is returned through the network interface the request was received from ($if_O$). In our example scenario, client B issues a request for content $x_1$. Once the request for $x_1$ reaches $R_3$, the C-FIB table of $R_3$ is updated to include $if_I=1$ and $if_O=2$ ($t_3$ in Table \ref{table:r2routing-table}). Then, at time $t_4$, the request for $x_1$ reaches $R_1$. Content chunk $x_1$ is found cached at $R_1$ whose interface 2 is marked as $if_O$ and the content is sent towards client B.

By introducing the $if_{TI}$ field in the C-FIB table we further realise off-path in-network caching \cite{scan}, as well as user-assisted in-network caching \cite{userAssistedCaching}, \cite{sit-networking-sourlas}. When a content chuck is not found in the local cache, the LIRA node sends the received request towards both the (permanent) incoming interface $if_I$ (as indicated by the name resolution process) and the temporary interface(s) $if_{TI}$. In our example, $R_1$ sends two requests for $x_1$ towards both the (permanent) $if_I$ 1 and the $if_{TI}$ 3 ($t_4$ in Table~\ref{table:r1routing-table}). Whichever of the two interfaces (1 or 3) starts receiving the requested data first is marked as the incoming interface for this content and the remaining (temporary) interfaces are pruned down. Pruning here can be realised through a negative ACK (NACK) packet which travels towards the source of the content. If $if_I$ answers first, the $if_{TI}$ is removed from the corresponding C-FIB table entry. Alternative strategies can be applied here, by selectively forwarding a request to one or more of the available interfaces \eg always forwarding only towards an off-path cache, since requests are always routable to the CP at the IP layer.

Finally, at time $t_5$ when $x_1$ transfer completes (from either the local, or a remote cache), interface 2 is added to the list of temporary incoming interfaces ($if_{TI}$) at $R_1$, since $x_1$ can now be found this way too (similarly to $t_2$). The C-FIB table of $R_3$ is  also updated to include interface 2 as $if_{TI}$ (step $t_6$).

We note that in order to avoid routing loops in case no other device towards $if_{TI}$ (client A in our case) has the content cached, we discard requests (for items in the C-FIB table) that come in through its marked $if_{TI}$. This is done because any LIRA node towards the $if_{TI}$ (client A in this case) will forward the request based on its IP address (carried at the IP layer and always pointing towards the permanent content source, hence through $R1$ in our example) if it finds no entry in its C-FIB table for the requested content.
In turn, upon receipt of the request, $R1$ will send the request back towards the same direction (towards client A here), since it still has got the related entry in its C-FIB table. This will result in the request travelling back and forth creating an endless routing loop.

\subsection{Multicast}\label{multicast}

Multicast support is enabled through the use of the $if_O$ and $mIP$ fields of the C-FIB table. As described above, during the chunk transfer, the network interface of the LIRA node where the incoming data is forwarded towards is marked in the $if_O$ field. This $if_O$ entry enables the LIRA node to suppress any subsequent request for the same content chunk by adding an extra outgoing interface to its C-FIB. This is similar to the PIT functionality in CCN \cite{ccn}.\footnote{Effectively, the C-FIB collapses both the CCN FIB and PIT in one table.}
Note that in all above steps the IP address (at the IP layer) of request packets has been pointing to the CP and of content chunks to the corresponding clients. However, in order to realise multicast transmission in this case (\ie avoid sending a second request for the same chunk towards the same direction), the LIRA node that suppresses subsequent requests needs to keep the extra IP address of the clients that generated the requests. We deal with this situation through the ``multicast IP" ($mIP$) field in the C-FIB table. When data arrives at the branching LIRA node, it gets forwarded to all $if_O$ interfaces. The $mIP$ entries are used at the IP layer to allow for the delivery of the duplicated data to the requesting recipients.

Note however that multicast forks further down the path are handled locally. In our example, if an additional client C attaches to $R_3$ and requests for $x_2$ during the multicast session, its request will be suppressed by $R_3$ which will also store client C's IP address in the corresponding $mIP$ field. $R_1$ will not be aware of client C's existence and $R_3$ is responsible for duplicating data for this client. Thus, the $mIP$ state load is distributed to the participating LIRA nodes avoiding the overloading of nodes closer to the root of the multicast tree.

In our example network, client A issues a request for content $x_2$. The C-FIB table at $R_1$ marks  $if_I = 1$ and $if_O = 3$ for \texttt{cID} $x_2$ (step $t_7$). Before the transfer of $x_2$ towards A completes through $R_1$ client B issues a request for $x_2$, which goes through $R_3$ and reaches $R_1$. $R_3$ updates its C-FIB table by putting $if_I = 1$ and $if_O = 2$ (step $t_8$). $R_1$ does not forward this request further; instead it adds interface 2 to the $if_O$ field of $x_2$ and also stores the IP address of client B (taken from the corresponding IP layer field) in the $mIP$ field (step $t_9$).

When $x_2$ arrives at $R_1$ (step $t_9$) it is forwarded towards client A through $if_O = 3$, but it is also replicated and forwarded towards client B, through $if_O = 2$, using $mIP$ as the destination IP address. When the chunk $x_2$ transfer completes, router $R_1$ moves interfaces 2 and 3 and $R_3$ moves interface 2 to the $if_{TI}$ field (step $t_{10}$ - Table~\ref{table:r1routing-table} and~\ref{table:r2routing-table}).
\\[0.2cm]
Note that the C-FIB table introduced here, incorporates the functionality of both the PIT and the FIB tables of CCN. For as long as the chunk transfer goes on and hence, the $if_O$ field is filled (and the $if_{TI}$ field is empty - $t_1, t_7$ and $t_9$ in $R_1$'s C-FIB, see Table~\ref{table:r1routing-table}), the C-FIB table represents the PIT table of CCN/NDN. That is, based on this state, LIRA nodes are able to collapse/suppress subsequent requests for content already requested (or under transmission) and realise multicast. When the chunk transfer completes and the entry in $if_O$ is moved to the $if_{TI}$ field ($t_2, t_4, t_5$ and $t_{10}$ in Table~\ref{table:r1routing-table}), then the C-FIB table reflects the FIB table of CCN/NDN. As mentioned above, however, the C-FIB table acts as a cache for recently served content and hence, it does not need to keep huge amounts of state information in the FIB part of the C-FIB. We discuss and evaluate both parts of the C-FIB table later in Section~\ref{eval}.


\makeatletter{}
\section{Performance evaluation}\label{eval}

It is generally not common to evaluate a network architecture merely in quantitative terms, given that the contribution of such studies comes mainly at a conceptual level.
In our case, the contribution of the LIRA architecture comes mainly in terms of incremental deployment with backward compatibility guarantees.
At the same time, however, LIRA can achieve all the quantifiable benefits of an ICN mode of operation.

To provide a thorough performance evaluation, we analyse conceptual and qualitative gains in Sec. \ref{qualitative-eval} as well as quantitative gains in Sec. \ref{quantitative-eval}.
The quantitative evaluation focuses on the deployment of the LIRA concept from the operators perspective. 
In particular, given a fixed monetary budget that the operator is prepared to spend in order to deploy LIRA, we assess the best strategies of investing the capital in terms of extra equipment, which in our case translates to cache memory and C-FIB tables. 
We also demonstrate and quantify the benefits brought by LIRA to CPs, with a particular focus on cache purging and the control over the freshness of the cached content. 

\subsection{Qualitative Evaluation}
\label{qualitative-eval}

\textbf{Name resolution:} by handing control of the name resolution process to CPs, LIRA avoids the need for either the deployment of a costly name resolution infrastructure, or the investment on in-network resources for the support of line-speed name resolution. 
The operation of the C-FIB as a cache for names/\texttt{cID}s is similar to \cite{conet-comnet}.
However, LIRA does not necessitate the use of an explicit off-path name resolution mechanism, as it rather falls back to IP, in a backward compatible manner. 
At the same time, by following a backwards compatible HTTP-supported name resolution mechanism, LIRA presents a complete interface for the interaction of end-hosts with an information-centric network. 
To the best of our knowledge, no exact mechanism has been proposed for the discovery (\ie not only resolution) of content names in alternative ICN architectural proposals.

\textbf{Control of content access:}
LIRA enables CPs to directly monitor and control the access of end users to their content. 
Route-by-name approaches such as \cite{ccn} fail to provide such support. 
CPs would be reluctant to accept transparent access to their content, thus dis-incentivising the adoption of such an approach to ICN by ISPs.
Lookup-by-name approaches, on the other hand, such as \cite{netinfnrs} and \cite{pursuitnrs}, enable this type of control, by decoupling name resolution from forwarding. 
However, this comes at the cost of additional name resolution infrastructure and directly places the content access information in the hands of ISPs; in turn, this introduces the burden of new (business and technical) interfaces between all CPs and all ISPs at a global scale.

\textbf{Mobility:} although the issue of mobility in case of LIRA requires further investigation and at first sight it 
might seem that LIRA cannot deal with mobility efficiently, due to its dependence on IP, we note the 
following: upon a content request, the CP or CDN is sending back to the client the \texttt{cID}s of the 
next few chunks, \ie not just the next one. That said, the clients operate based on IP-agnostic \texttt{cID}s. Therefore, client 
mobility can be natively supported, as clients request for content based on identifiers (in combination to the IP address at the IP layer). Source mobility, on the other hand, is an issue that requires further investigation as is the case with all ICN architectural proposals.

\textbf{Security:}
by supporting self-certifying \texttt{cIDs}, LIRA secures the content itself rather than the communication channel, similarly to other ICN architectures \eg \cite{ccn}.

\textbf{Implementation:}
the proposed LIRA functionalities can be deployed on nodes with only firmware updates without the need for hardware replacement or upgrade.
In fact, by relying on IP forwarding as a fallback in case of C-FIB misses, LIRA will never result in un-routable requests/content even if deployed on just a few nodes and with minimal memory.
This is in stark contrast with previous ICN proposals like CCN and NDN which require well-dimensioned FIB and PIT structures to operate correctly and at line speed.
C-FIB can be loaded in DRAM, which has been shown to be able to support line-speed per-packet lookups \cite{cuckooswitch-conext13}, \cite{caesar-ancs14}, is inexpensive and abundant on modern routers based on either network processors or general purpose processors.
It is in fact common to have at least a few GBs of spare DRAM on modern routers.
Since the binary code implementing LIRA functionalities is likely to require negligible space, all available DRAM can be used for C-FIB and caching space. C-FIB entries in particular have very low memory requirements. 
In fact, even assuming that (i) the C-FIB is implemented using a hash-table with a load factor of $0.5$ and with a circular queue for replacement and (ii) the unfavourable case that LIRA chunks are named using SHA-512 hashes and next hops information are coded on 2B, it is still possible to store over 15 million C-FIB entries per GB of DRAM.
This makes C-FIBs and more generally the LIRA node architecture easy to incrementally deploy on today's routers.

\subsection{Quantitative Evaluation}
\label{quantitative-eval}

As mentioned earlier, the objective of this section is to evaluate the best possible way to invest in 
deploying the LIRA concept, from the operator's perspective. That said, we initially evaluate the main 
concepts of our proposal with regard to their projected gains in terms of cache hits. 
Although LIRA is far from a caching-specific architecture, caching is: \textit{i)} the only 
straightforward quantitatively measurable aspect of an ICN architecture, and most importantly, \textit{ii)} 
the main feature that requires investment from network operators. For these reasons and without by any 
means underestimating the gains from the above-mentioned qualitative benefits of the LIRA 
architecture, in this section, we focus on the evaluation of the main concepts included in LIRA as seen 
from an in-network caching perspective.

We use Icarus \cite{icarus} to evaluate the performance of various aspects of our proposed framework based on real ISP topologies from the RocketFuel dataset \cite{rocketfuel-sigcomm02} and synthetic workloads \cite{globetraff}. Due to space limitations, we omit evaluation of the multicast functionality offered by LIRA, since the related performance benefit is straightforward. Moreover, we only show results for the Telstra and Abovenet topologies, though we report that we obtain similar results with other topologies as well. We make the code, documentation and data required to reproduce our results publicly available.\footnote{\url{http://www.ee.ucl.ac.uk/~lsaino/software/lira}}

\subsubsection{Efficiency of C-FIB and Deployment Strategies}

LIRA nodes can have a content cache or a C-FIB table, or both.
Given a fixed total cache and C-FIB capacity budget, in this section, we identify the best possible combination of cache and C-FIB deployment along two dimensions: i) deployment strategies and ii) caching strategies.
We attempt to capture the interaction dynamics between nodes that cache content and nodes that can route to this content, in a location-independent manner, \ie through C-FIB table entries.
\textit{Our first objective is to investigate the effectiveness of C-FIB table entries in mapping the content cached in neighbour nodes. Our second objective is to see how C-FIB table entries eventually translate to cache hits.}

Modelling the performance of a network of caches is known to be complex \cite{anet-infocom13}, \cite{complexity-infocom15}. As a result, it is extremely difficult to formulate optimal cache placement algorithms which are also robust to realistic traffic variations.
Arguably, the complexity of the optimal cache placement problem is another obstacle hindering ICN deployments.
Therefore, motivated by practical reasons, we propose four simple content cache and C-FIB placement algorithms and show that they are sufficient to provide tangible performance gains even with partial deployments.
To deploy caches and C-FIBs, we rank nodes according to their betweenness centrality (\ie the amount of traffic traversing them following shortest path routing \cite{clam-comcom}) and deploy LIRA functionality using the following strategies:
\\
\textit{(i)} Cache in top 50\% high centrality nodes, C-FIB table in all nodes: $\left(C_{H}, F_{A} \right)$.\\ 
\textit{(ii)} Cache in top 50\% high centrality nodes, C-FIB table in top 50\% high centrality nodes: $\left(C_{H}, F_{H} \right)$.\\
\textit{(iii)} Cache in all nodes, C-FIB table in all nodes: $\left(C_{A}, F_{A} \right)$.\\
\textit{(iv)} Cache in all nodes, C-FIB table in top 50\% high centrality nodes: $\left(C_{A}, F_{H} \right)$.

\begin{figure}[!t]
\centering
\subfloat[Telstra]{\label{freshness-1221}\includegraphics[width=0.23\textwidth]{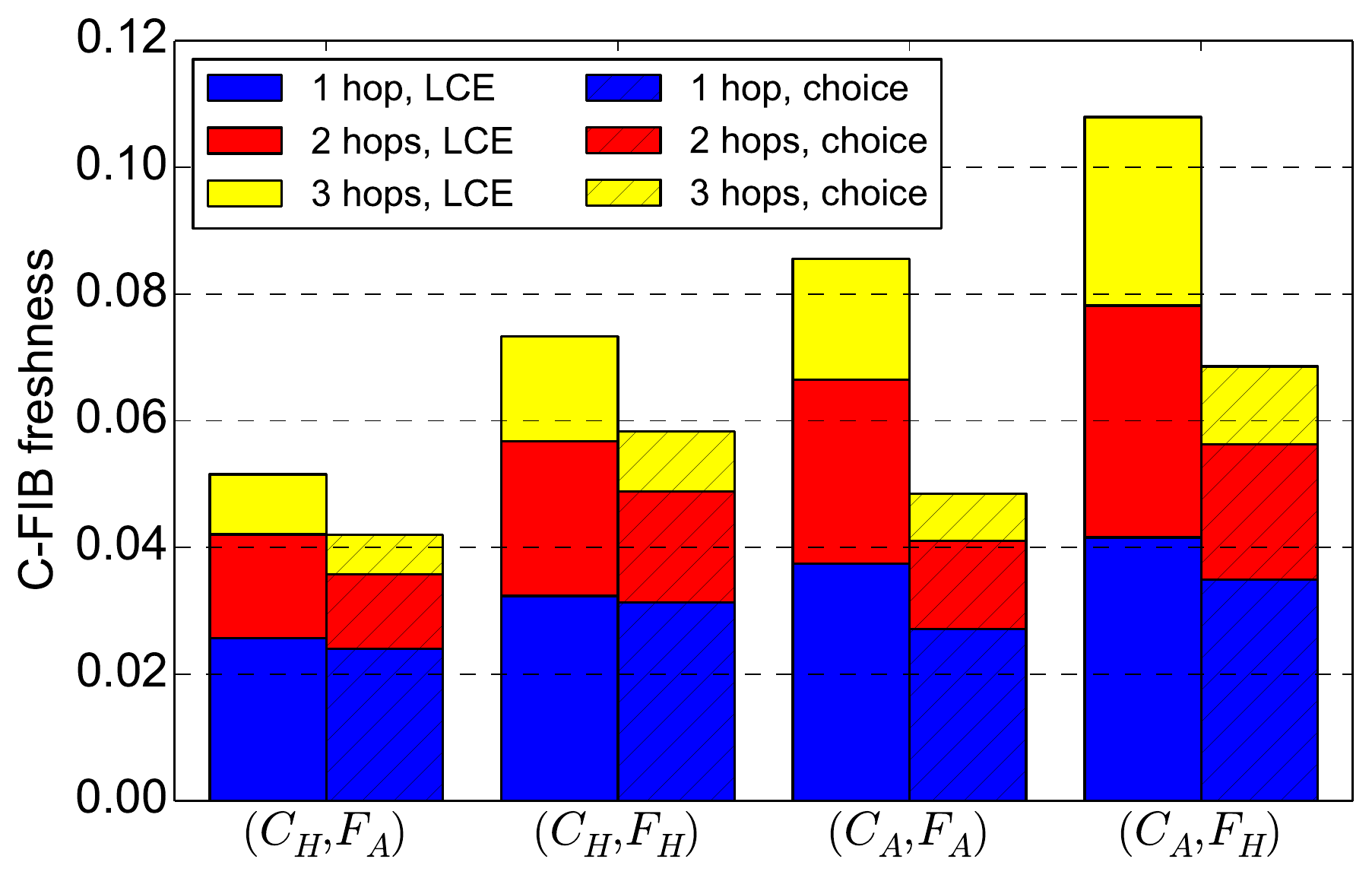}} 
\subfloat[Abovenet]{\label{freshness-1755} \includegraphics[width=0.23\textwidth]{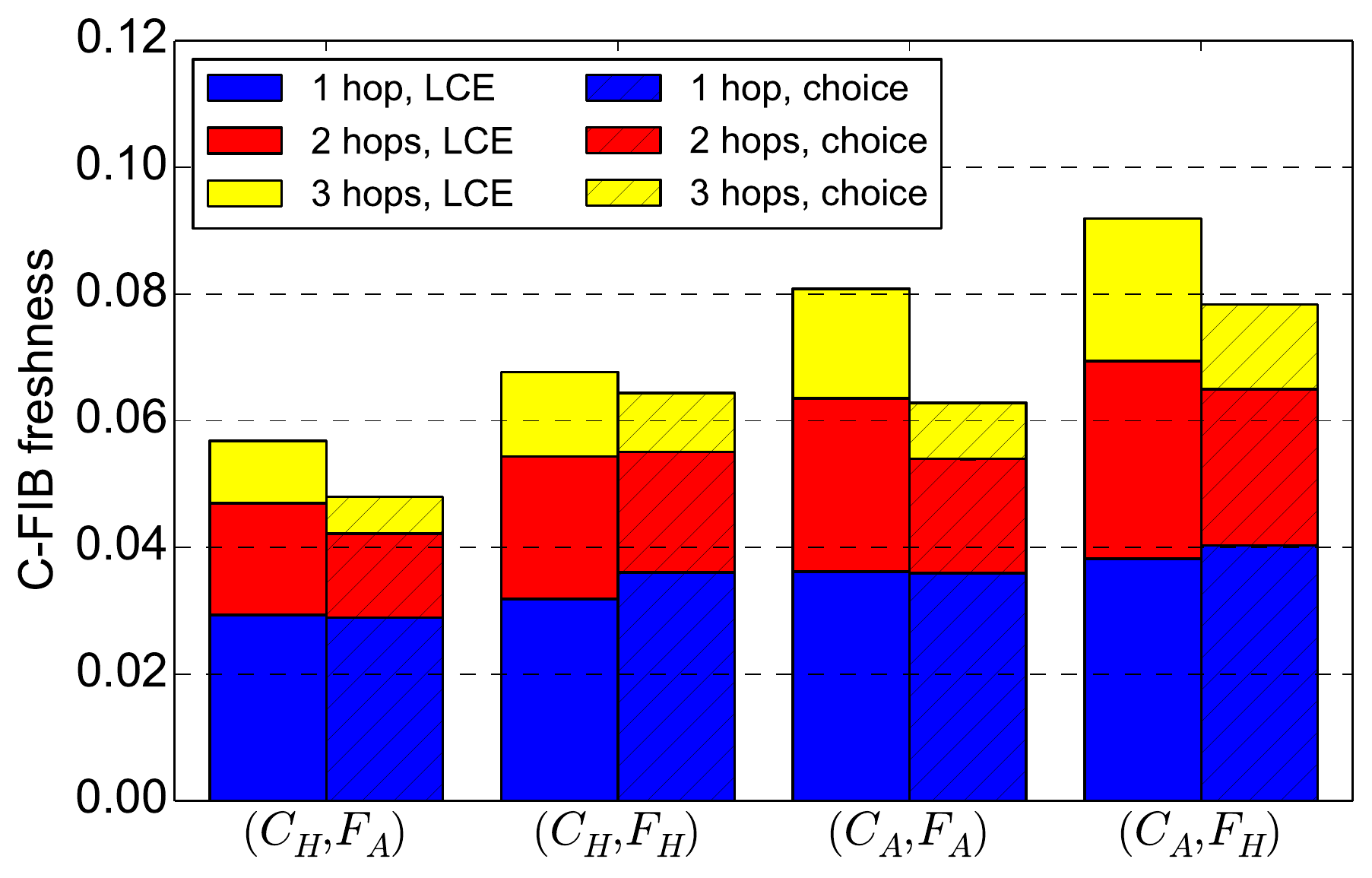}}
    	\caption{C-FIB Freshness vs. Deployment Strategies and Caching Strategies, Zipf $\alpha=0.8$}
    	\label{C-FIB-freshness}
\end{figure}

\begin{figure}[!t]
\centering
\subfloat[Telstra]{\label{cachehits-1221}\includegraphics[width=0.23\textwidth]{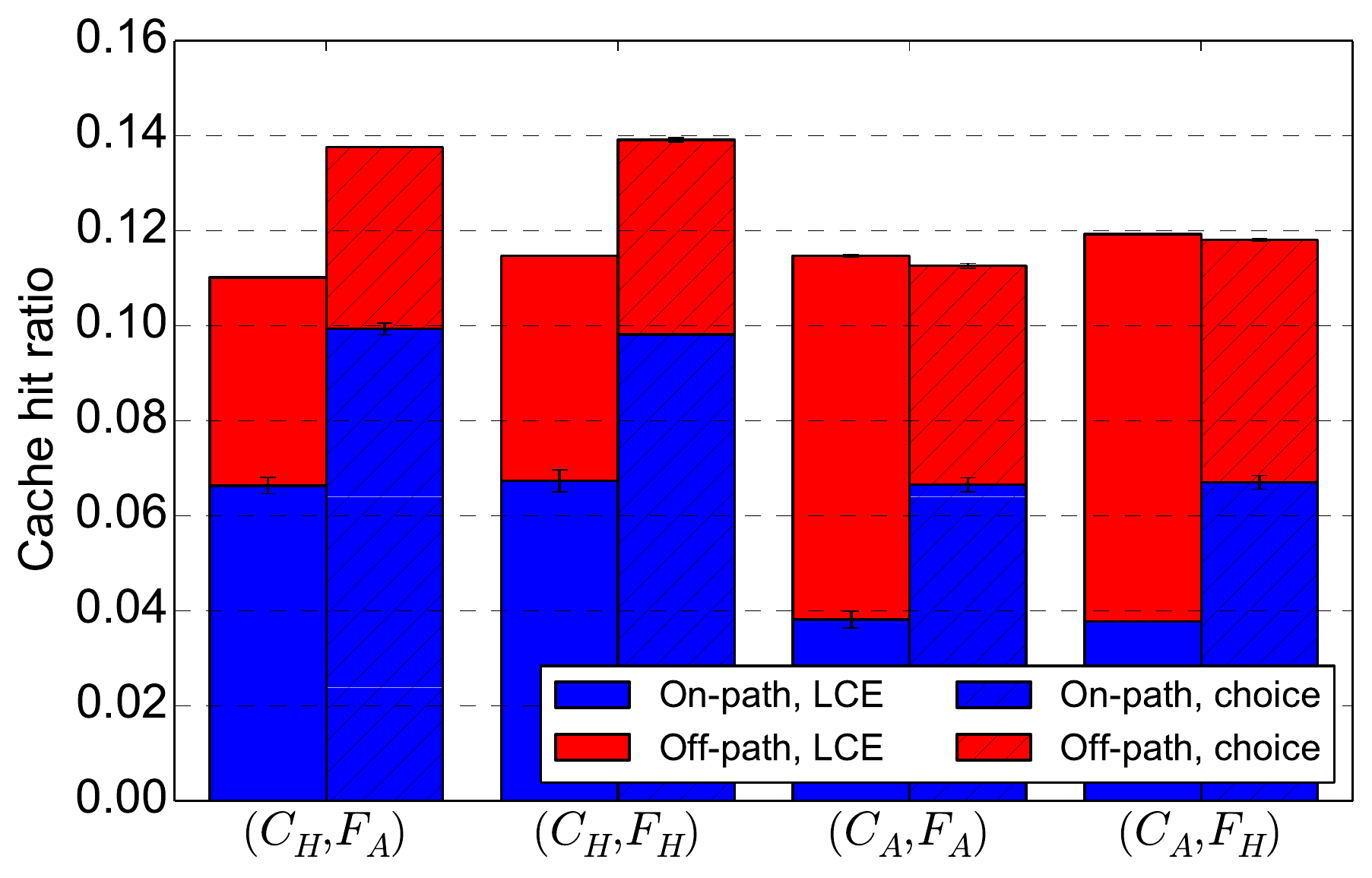}} 
\subfloat[Abovenet]{\label{cachehits-1755} \includegraphics[width=0.23\textwidth]{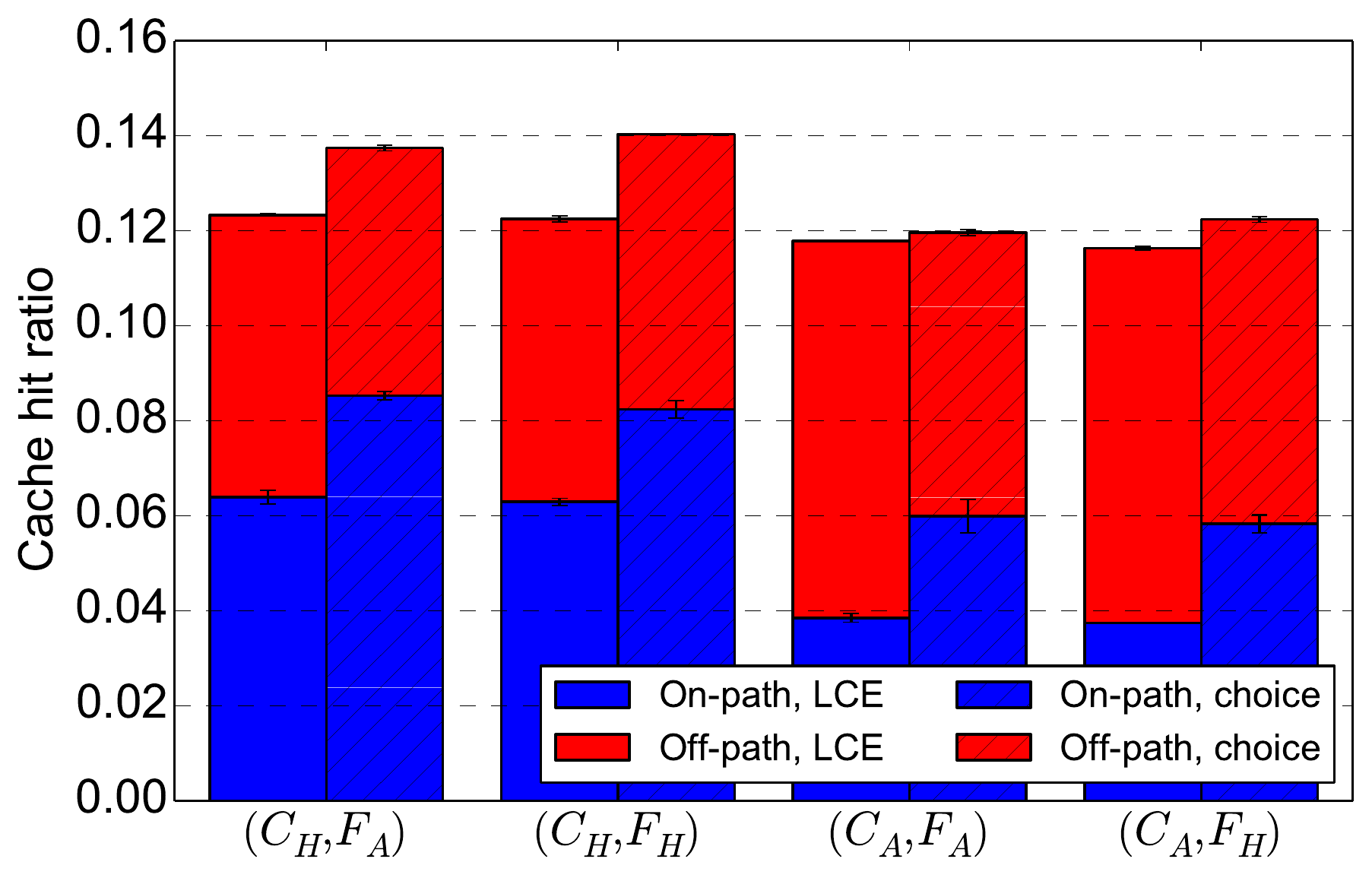}}
    	\caption{Cache Hit Ratio vs. Deployment Strategies and Caching Strategies, Zipf $\alpha=0.8$}
    	\label{deployment-strategies}
\end{figure}

We run simulations and measure the mean \textit{C-FIB freshness}, which we define as the ratio of entries stored in C-FIB tables which can correctly route to a copy of a content stored in a nearby cache.
This metric captures how well the entries of the C-FIB tables deployed in the network reflect the current state of nearby caches.
We further characterise the correct C-FIB entries by the hop distance to the LIRA node that caches the corresponding content.
Note that in all cases, and regardless of the deployment strategy, the ratio of C-FIB table to cache entries is fixed (see next subsection for the evaluation of this ratio).
As a result, C-FIB tables in fewer nodes (than those that deploy caches) keep more entries to match the number of cache slots (and vice versa).

We also analyse the results under different caching strategies: \textit{Leave Copy Everywhere} (LCE), according to which a copy of a content is stored in every cache traversed and \textit{random choice}, according to which a content is stored only in one randomly selected caching node along the delivery path.
The rationale behind our choice is to evaluate deployment and caching performance under varying \textit{caching redundancy} \cite{probcache-tpds}. Our results are shown in Figs.~\ref{C-FIB-freshness} and \ref{deployment-strategies}.
\\[0.2cm]
\textbf{C-FIB Efficiency.} First of all, it is important to highlight the fact that the C-FIB table entries depict precisely the state of neighbour caches. This is proved by the fact that the \textit{freshness ratio} in Fig.~\ref{C-FIB-freshness} directly translates to off-path cache gain in Fig. \ref{deployment-strategies}: for instance, the freshness result in case of $\left(C_{H}, F_{A} \right)$ in Fig.~\ref{freshness-1221} indicates that 5\% of entries in the C-FIB table can correctly route to the content in neighbour caches. In turn, in Fig.~\ref{cachehits-1221}, the gain from off-path caching (red, top part of bar) is 4.5\%. This is an important result that highlights the effectiveness of the C-FIB table in keeping an accurate record of the state of nearby caches (\ie up to 3 hops away in our evaluation).
\\[0.2cm]
\textbf{Deployment Strategy.} In terms of C-FIB freshness, deploying smaller caches over more/all nodes, \ie $\left(C_{A}, F_{*} \right)$, seems to be more effective in capturing the state of caches from the C-FIB tables (\ie higher freshness in Fig.~\ref{C-FIB-freshness}). This is explained by the fact that the ``monitoring and mapping" mechanism provided by the C-FIB table has got a wider view of the neighbourhood and can therefore, find more content items locally. This also translates to more off-path cache hits in Fig.~\ref{deployment-strategies} for $\left(C_{A}, F_{*} \right)$.

Out of the four deployment strategies under consideration here, $\left(C_{H}, F_{H} \right)$ and $\left(C_{H}, F_{A} \right)$ consistently perform best in terms of cache hits (in Fig.~\ref{deployment-strategies}). This is irrespective of the freshness result, which shows that freshness improves when caches are deployed over all nodes (\ie $\left(C_{A}, F_{*} \right)$). In other words, it is better to have fewer but bigger caches placed in high centrality nodes (as also shown in \cite{clam-comcom}), rather than having smaller caches deployed in all nodes of the network. 
\\[0.2cm]
\textbf{Caching Strategy.}
As expected, in terms of cache hits, \textit{choice} always performs best, for all topologies and for all deployment strategies, as a result of its reduced caching redundancy. Similar results have been reported before in \cite{probcache-tpds}. LCE on the other hand, performs roughly the same across all deployment strategies. Note that the LCE result in Fig.~\ref{deployment-strategies} effectively reveals the performance of the CCN/NDN architecture. Due to space limitations, we do not present a full-fledged comparison between the architectures, but Fig.~\ref{deployment-strategies} reveals very well the cache-related performance of CCN/NDN.

\subsubsection{Memory Requirements and Scalability}

\begin{figure}[h]
\centering
\subfloat[LCE]{\label{hits-abovenet-lce}\includegraphics[trim= 0 5 0 20, width=0.5\linewidth]{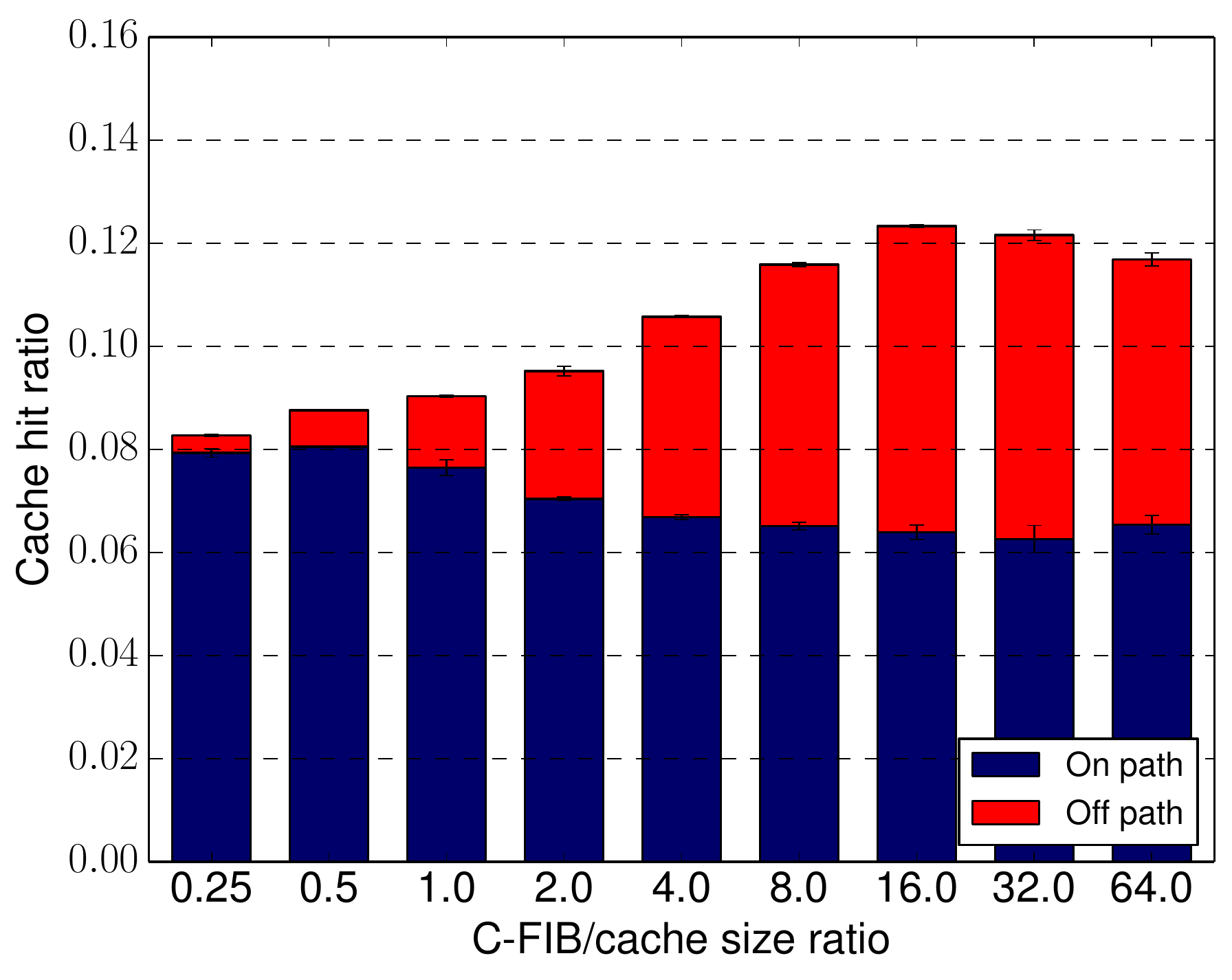}}
\subfloat[Random Choice]{\label{hits-abovenet-choice} \includegraphics[trim= 0 5 0 20, width=0.5\linewidth]{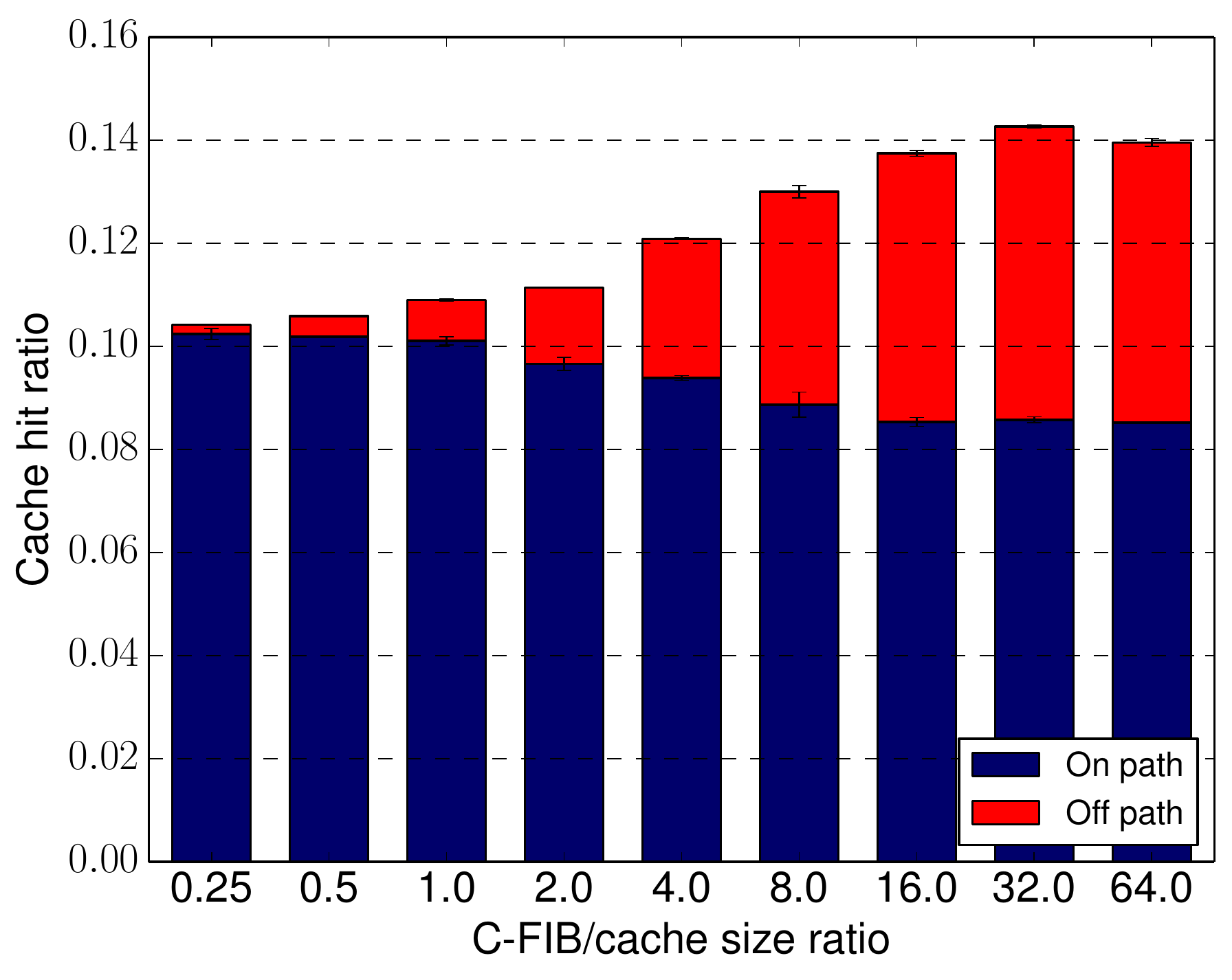}}\vspace{-0.1cm}
    	\caption{Cache hit ratio vs C-FIB size, Abovenet}
    	\label{hits-vs-C-FIB}
\end{figure}

We next quantify the performance benefit of off-path, C-FIB-routed, caching for various values of the C-FIB-to-cache size ratio (expressed in number of entries) and for the $\left(C_{H}, F_{A} \right)$ strategy (Fig.~\ref{hits-vs-C-FIB}).

Considering the overall cache hit ratio (both on- and off-path), we see a considerable increase when moving from a ratio value of 0.25 to a ratio value of 16, due to C-FIB routing 
redirections. The results are similar for a ratio equal to 32, but the gain in this case is marginal. 
Therefore, given that larger memory is required in order to deploy C-FIB tables 32 times bigger than the entries in the respective caches, we conclude that a value of 16 is optimal.
Although in absolute values, off-path, 
C-FIB-based, caching contributes less than on-path caching, the gain is still far from negligible (\ie it can reach up to 50\% in Fig. \ref{hits-vs-C-FIB}).
We report that in our simulations, the gain from off-path caching can reach 100\%, effectively doubling the gain from on-path caching.

Finally, it is interesting to note the slight decrease of on-path cache hit ratio as the C-FIB-to-cache size ratio increases.
This is attributed to cases where a content request encounters a C-FIB table entry and gets diverted to an off-path cache, before it actually hits an existing on-path cache.
In this case, and given that the C-FIB table entry is found earlier in the path, we report that the delay to deliver the content back to the user is even shorter than finding the requested item in an on-path cache.
This is especially so in case of LCE caching, where due to increased caching redundancy, a copy of a content has good chances of being found along the shortest path.

\subsubsection{Control of content to CPs}

One of the departing points in the design of the LIRA architecture is the direct control of content by the CPs/CDNs, as discussed earlier. We identify two main features that give direct control of the content to the CP or CDN. The first one is the control of access logging. In LIRA this is accomplished by the content provider-controlled name resolution, where clients need to get the up-to-date \texttt{cID} from the content provider. This requires an extra RTT to get to the CP or CDN. We remind that according to our discussion in Section~\ref{transitioning-interval}, CPs/CDNs send more than one \texttt{cID} to the client, therefore, the journey to the CP/CDN happens rarely during the data transfer, or even only once in case of small files (\eg web). We assume this extra RTT to incur only a tiny performance penalty compared to alternative proposals that do not necessarily require this extra roundtrip.

The second feature that provides control of published content to CPs is the ability to actively perform cache purging. As described in Section~\ref{cc:en}, when CPs change the \texttt{cID} of a content item, previously cached items no longer get hits from new requests and eventually get evicted (denoted as \textit{LIRA w/o replacement}). 
Taking a step further, we consider an extended version of this mechanism, where data packets explicitly indicate the \texttt{cID} values of the content items that should be immediately evicted from encountered caches (denoted as \textit{LIRA w/ replacement}).

\begin{figure}[!h]
\centering
\subfloat{\label{stale-hits-stationary}\includegraphics[width=\linewidth]{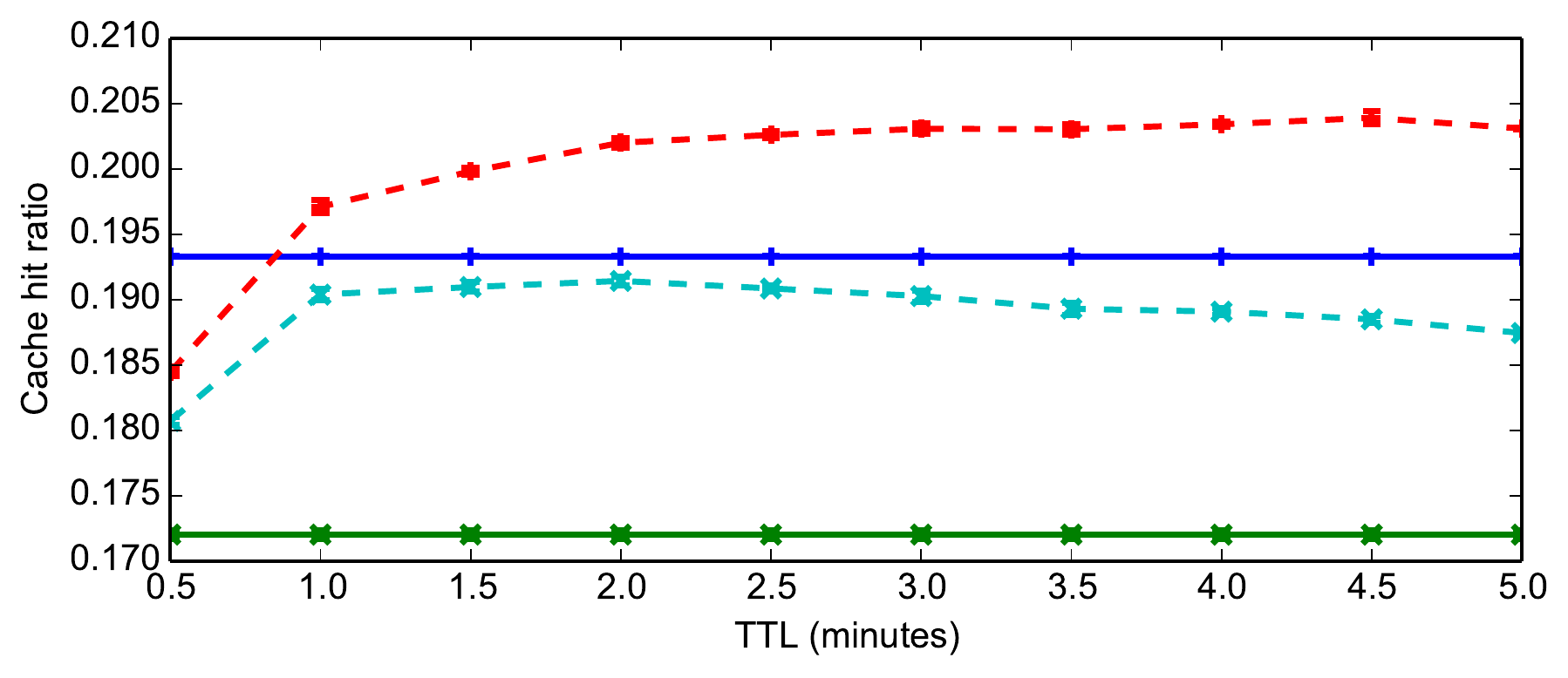}} \vspace{-0.1cm}
\subfloat{\includegraphics[width=0.8\linewidth]{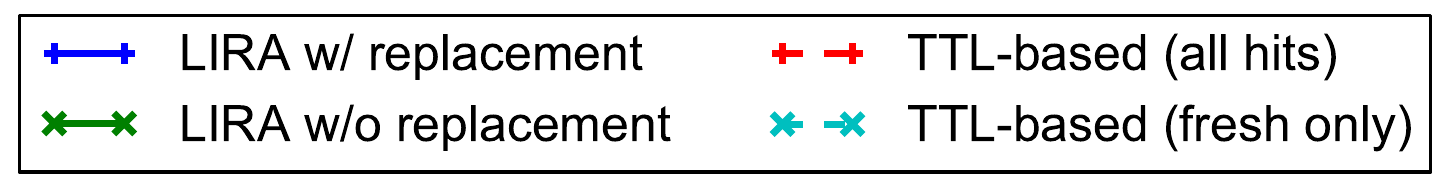}}\vspace{-0.2cm}
    	\caption{Delivery of Stale Content}
    	\label{fig:purging-cache-hit-ratio}
\end{figure}

Figure~\ref{fig:purging-cache-hit-ratio} shows the cache hit ratio of the above mechanisms along with 
that of a simple TTL-based mechanism, where any cache hit returns the content to the client, even if this content is stale (\textit{TTL-based (all hits)}) for various TTL values. In Fig.~\ref{fig:purging-cache-hit-ratio}, we see that the cache hit ratio in the 
(\textit{TTL-based (all hits)}) case increases with the value of TTL, since content remains longer in the 
cache. However, this also means that the corresponding cache hits result in the reception of stale 
content. Fig.~\ref{fig:purging-cache-hit-ratio} also shows the cache hit ratio for fresh only content 
(\textit{TTL-based (fresh only)}), which initially increases, but then steadily drops as a 
result of high TTL values that increase stale cached content.

The \textit{LIRA w/ replacement} mechanism performs considerably better than \textit{LIRA w/o 
replacement}, as it immediately frees the caching space from unnecessary stale content, and better 
than its \textit{TTL-based (fresh only)} counterpart.

It must be noted that the \textit{TTL-based (fresh only)} ratio is only provided as a benchmark, 
as TTL-based mechanisms cannot avoid serving stale content. 
On the other hand, LIRA provides a precise mechanism to avoid serving stale cached content altogether.

\subsubsection{Incremental Deployment}

We proceed to evaluate the last of the design targets behind LIRA, that of incremental deployability. To assess the performance gain of incrementally deploying the LIRA architecture, we begin by progressively adding C-FIB tables starting from the highest centrality nodes. We evaluate the performance in terms of cache hits in case of caches deployed in 25\%, 50\%, 75\% and 100\% of the nodes, starting from the highest centrality ones.

\begin{figure}[h]
\centering
\subfloat[Telstra (108 nodes)]{\label{hits-abovenet-lce}\includegraphics[trim= 0 5 0 20, width=0.5\linewidth]{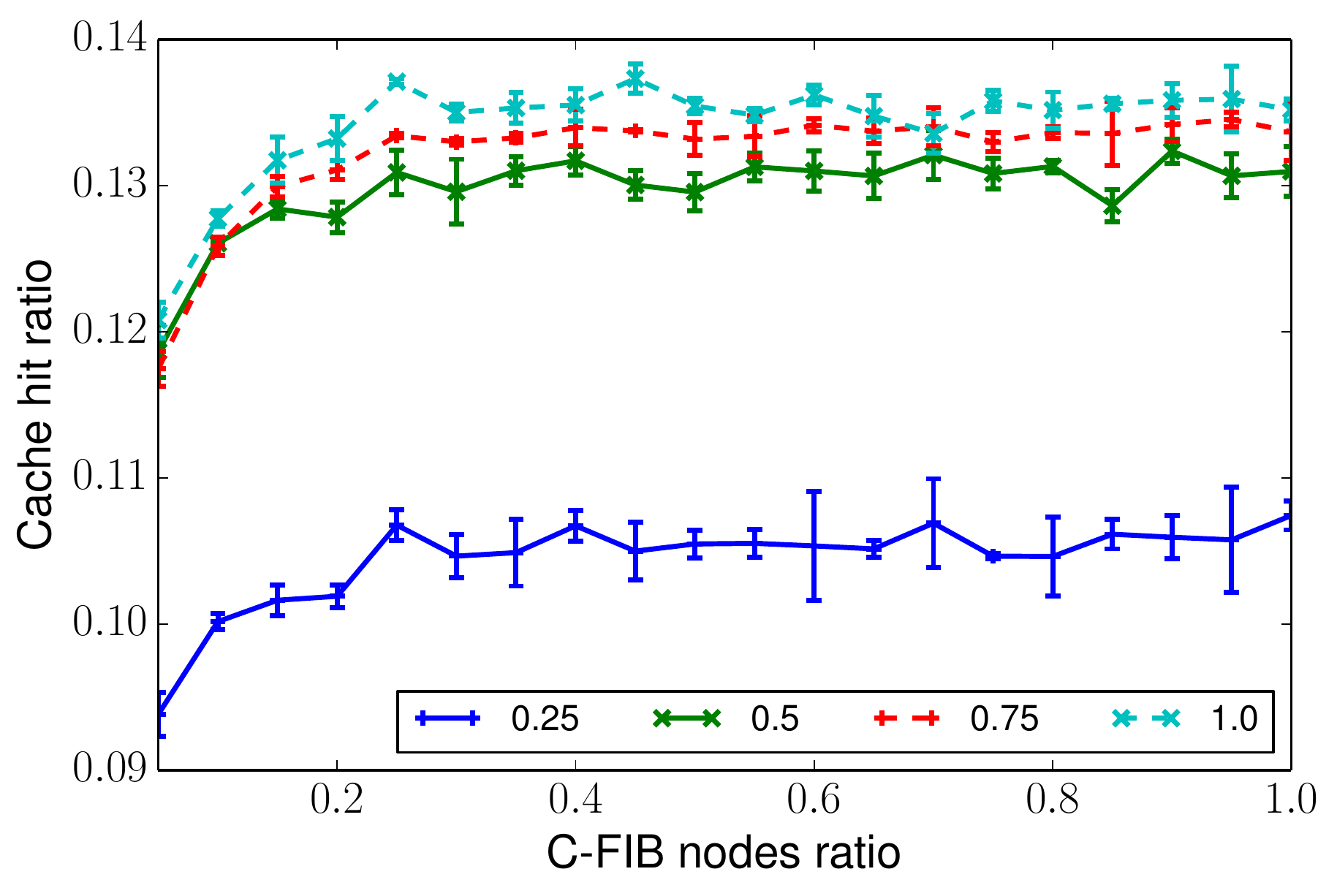}}
\subfloat[Abovenet (141 nodes)]{\label{hits-abovenet-choice} \includegraphics[trim= 0 5 0 20, width=0.5\linewidth]{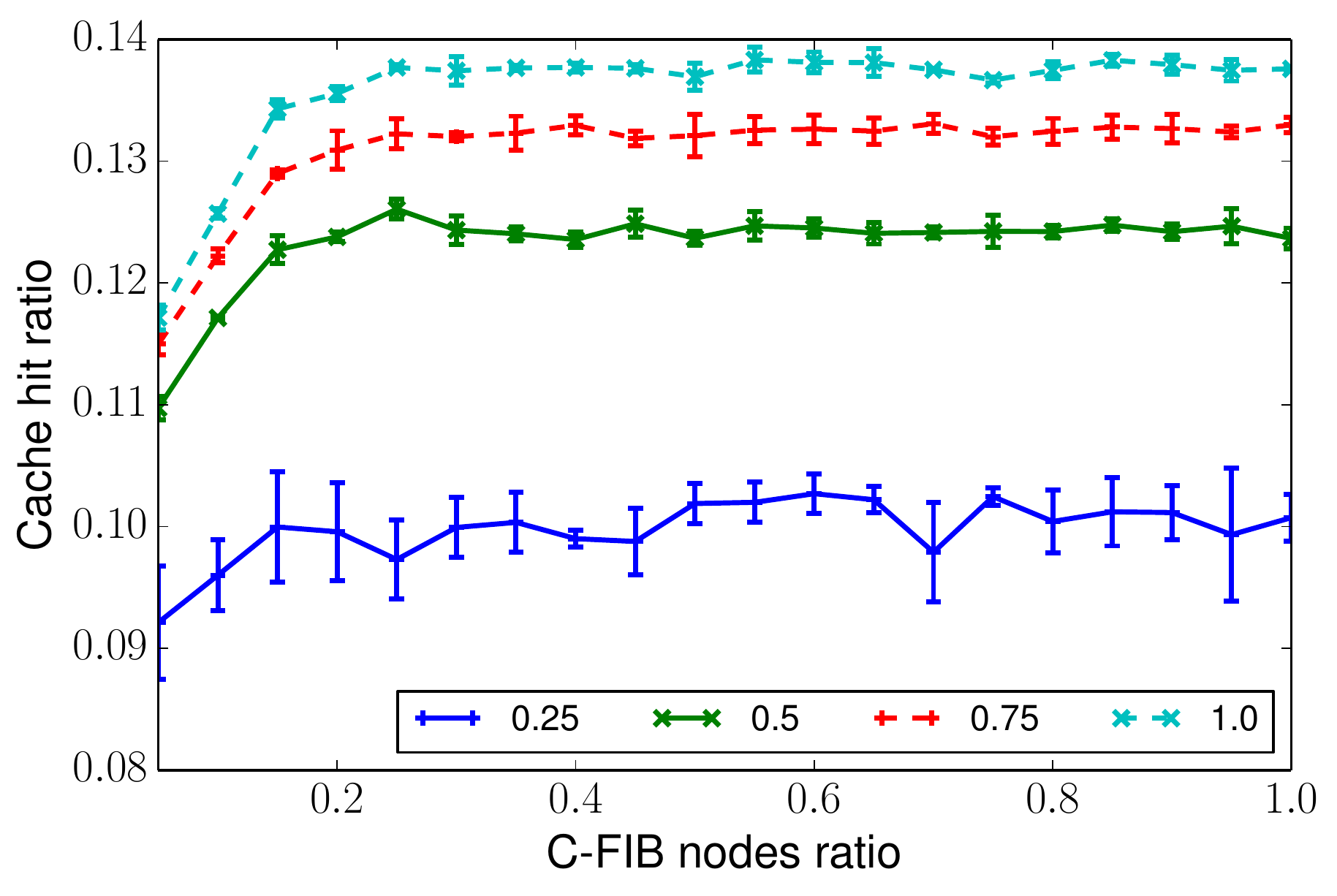}}\vspace{-0.1cm}
    	\caption{Incemental Deployment}
    	\label{fig:incremental-deployment}
\end{figure}

We observe in Fig.~\ref{fig:incremental-deployment} that performance stabilises with the C-FIB table present in 20-30\% of the nodes. C-FIB in less than 20\% results in suboptimal performance, but performance does not increase considerably if we continue adding C-FIB to more nodes. In terms of caches, a 25\% deployment rate results in poor performance, while the performance does not improve considerably when caches are deployed in more than 50\% of nodes. The difference in performance between the 50\% and 100\% of nodes is in the area of 1\% improvement in terms of cache hit ratio for the two topologies shown here (Telstra and Abovenet).

We conclude that adding C-FIB to the top 20-30\% highest centrality nodes and caches to 50\%-75\% of highest centrality nodes achieves the full performance gain of the LIRA architecture. Although here we present results for Telstra and Abovenet topologies, our results are consistent along all six evaluated topologies of the RocketFuel dataset.

\makeatletter{}\section{Conclusions}\label{concl}

There is a constant trend towards extra ``flexibility" in communication networks, which started with the shift from (rigid) circuit-switching to (queuing-based) packet-switching \cite{kurose-icn}. We see location-independent, information-centric networking as the natural next step towards \textit{``content switching"}. To move towards this direction, however, the research community needs to take into account the interests of the main Internet market players, as well as those of users.

We argue that ICN research so far has focused on designing conceptually sound and scalable name-based routing architectures, but largely ignored any incentives (provided through those architectures) to adopt the ICN technology. The interests of content providers and CDNs are largely different to those of ISPs and the shift to an ICN environment environment makes this difference even more pronounced. That said, unless a shift to an ICN environment takes into account the interests of both CPs/CDNs and ISPs, the incentives to adopt this technology will be limited.

In this paper we have taken these concerns into consideration and have designed an incrementally-deployable ICN architecture. The proposed architecture is based on the Location-Independent Routing Layer (LIRA) and directly involves the content provider in the name resolution process. Furthermore, ephemeral names give more power to the CPs/CDNs over the content they publish. Our evaluation shows that even with a limited number of nodes implementing the LIRA architecture, ISPs achieve a clear performance gain, while at the same time CPs/CDNs have full control of their content.

\section{Acknowledgments}

This work was supported by EPSRC UK (COMIT project) grant no. EP/K019589/1 and EU FP7/NICT (GreenICN project) grant no. (EU) 608518/(NICT)167.

The research leading to these results was funded by the EU-Japan initiative under European Commission FP7 grant agreement no. 608518 and NICT contract no. 167 (the GreenICN project) and by the UK Engineering and Physical Sciences Research Council (EPSRC) under grant no. EP/K019589/1 (COMIT).

\bibliographystyle{abbrv}

\end{document}